# How should AI knowledge be governed? Epistemic authority, structural transparency, and the case for open cognitive graphs

*Chao Li[a], Chunyi Zhao[b], Yuru Wang[a], Yi Hu[c]*

[a] *School of Information Science and Technology, Northeast Normal University, China*

[b] *Centre of Educational Design and Innovation, University of Otago, New Zealand*

[c] *School of Information Science and Engineering, Southeast University, China*

**Abstract**

Through widespread use in formative assessment and self-directed learning, educational AI systems exercise de facto epistemic authority. Unlike human educators, however, these systems are not embedded in institutional mechanisms of accountability, review, and correction, creating a structural governance challenge that cannot be resolved through application-level regulation or model transparency alone. This paper reconceptualizes educational AI as public educational cognitive infrastructure and argues that its governance must address the epistemic authority such systems exert. We propose the Open Cognitive Graph (OCG) as a technical interface that externalizes pedagogical structure in forms aligned with human educational reasoning. By explicitly representing concepts, prerequisite relations, misconceptions, and scaffolding, OCGs make the cognitive logic governing AI behaviour inspectable and revisable. Building on this foundation, we introduce the trunk-branch governance model, which organizes epistemic authority across layers of consensus and pluralism. A case study of a community-governed educational foundation model demonstrates how distributed expertise can be integrated through institutionalized processes of validation, correction, and propagation. The paper concludes by discussing implications for educational equity, AI policy, and sustainability. By shifting attention from access to governance conditions, the proposed framework offers a structural approach to aligning educational AI with democratic accountability and public responsibility.

**Contact** Chunyi Zhao, E-mail address: cccchunyi07@gmail.com

# 1. Introduction

Artificial intelligence (AI) has been widely applied in the field of education. In many existing studies, one of the most emphasized advantages of AI in teaching is its ability to provide timely feedback (Zhao, 2025a). Trained by large-scale data, AI integrates relevant knowledge in response to students' questions and produces conclusions in time(Al-Marzouqi et al., 2024). Over a long period of time, this capability has been understood as a factor that improves learning efficiency (Elbaz et al., 2024). Some researchers argue that AI may change future teaching models, and some even suggest that AI could partially replace the role of teachers (Chan & Tsi, 2024).

At the same time, doubts and concerns about AI remain widespread. These concerns include risks related to information security, inaccurate answers, weak logical coherence, the black-box nature of large models, hallucination phenomena, and the contamination of training databases with non-academic information (Floridi & Chiriatti, 2020; ;Đerić et al., 2025). The inherent limitations of large models, together with these concerns, have reduced trust in the use of AI in educational contexts (Bender et al., 2021). They have also created new difficulties for educational governance and have limited the extent to which AI can effectively serve education.

When students attempt to treat AI as a teacher and ask subject-related questions, AI often provides responses that remain at the level of surface knowledge and information (Li et al., 2025). These responses are generated through the integration of existing data in the model's database. After receiving such answers, students may feel that the interaction is similar to asking questions of a human teacher and obtaining responses. However, the underlying process through which these answers are produced is fundamentally different.

In the formative assessments, teachers can explain what the correct answer is and can also clarify why the student's reasoning deviated from an appropriate path (Shute, 2008). This judgment is grounded in the teacher's pedagogical experience. If the student disagrees with the result of the assessment, the teacher can indicate possible ways to appeal, such as consulting another teacher, reviewing textbooks, or referring to academic literature (Sadler, 1989). In this process, the teacher is both the executor of the judgment and the party responsible for it (Biesta, 2015). If the judgment is incorrect, it can be identified, questioned, and corrected.

When AI performs teacher-like functions such as diagnosing student understanding, planning learning pathways, and adjudicating correctness, no expert is able to examine the basis of these judgments due to the black-box structure of large models (Rudin, 2019). Intervention becomes impossible when judgments are incorrect. This situation is not simply a technical issue related to insufficient transparency (Doshi-Velez & Kim, 2017). It reflects a fundamental change in the structure of authority (Zhao, 2025b). A system that performs cognitive judgment functions operates under the label of a tool while remaining exempt from the accountability mechanisms that traditionally constrain human experts.

The issue of AI replacing teachers in exercising authority within learning-support dialogues has received limited attention in previous research. Most studies on AI-assisted learning and AI-assisted assessment focus on how instructional design and assessment design can help students improve AI literacy and use AI more responsibly (González-Calatayud et al., 2021). In these discussions, AI is still treated as a conventional tool that improves efficiency when compared with traditional tools. The expert status that AI has already acquired in practice is often overlooked.

A closer examination of what AI actually does reveals a different picture. AI judges whether a student's understanding is correct, determines the order in which concepts should be learned, and evaluates which explanations are suitable for a given student (Wayahdi & Zaki, 2025). These functions are expert functions. Before the emergence of AI, such judgments could only be made by human experts who were trained, responsible, and accountable.

Today, AI performs these expert functions at scale without inheriting corresponding accountability mechanisms. It claims the exemption associated with tools while exercising the authority of an agent. Due to the absence of accountability, errors caused by the limitations of large models cannot be effectively repaired. Even when a model is found to consistently mislead students on a specific concept, there is no mechanism that allows human educators to access the internal structure of the model and correct that specific cognitive pathway. In practice, society passively waits for AI companies to retrain entire models or attempts to guide students through courses that emphasize cautious use of AI. These remedial approaches are inefficient and often expose students to continued guidance based on incorrect information.

To address this problem, this study proposes a Trunk-Branch model based on the concept of the open cognitive graph. The model aims to define who has the authority to modify cognitive structures and how consensus on such modifications can be achieved. Drawing on governance practices from open-source software communities, this study seeks to establish a balance between professional educational authority and structured community participation.

**2. Conceptual framework: public educational cognitive infrastructure**

2.1 The epistemic authority of AI and its risks

In traditional educational systems, assessment is supported by institutional arrangements that confer epistemic authority (Jäger, 2016). Disciplinary knowledge is produced through peer review, community consensus is formed through sustained academic debate, and teachers acquire the authority to make pedagogical judgments through professional training and certification. Within this system, specific evaluative judgments are subject to appeal, review, and revision. It is this institutionalized process that grants legitimacy to educational assessment and explains why such judgments are widely accepted.

When artificial intelligence systems, particularly large language models, participate in educational assessment, this foundation of authority is altered. From a technical perspective, large language models are trained on corpora that lack consistent and transparent quality control. Their training processes are opaque, their internal parameters are not interpretable, and their outputs cannot be reliably traced to identifiable sources of knowledge. As a result, assessments generated by such systems are not embedded in the established mechanisms through which epistemic authority is traditionally produced and maintained in education.

At the same time, a central tension arises from the functional role these systems perform in practice. Although AI has not been widely adopted for formal, outcome based assessment in which it would directly replace teachers in issuing final evaluations of academic performance, it has become increasingly prevalent in informal contexts, particularly in students' self-assessment or self-reflection practices (Jelodari et al., 2023). Students frequently rely on AI systems to judge whether their understanding is adequate, to determine subsequent learning steps, and to estimate their level of mastery. In these contexts, AI assisted personal assessment performs functions that are equivalent to those traditionally carried out by human educational experts.

Through this widespread use, AI systems acquire a form of de facto epistemic authority, insofar as they come to function as trusted sources that shape beliefs and judgments despite lacking formal institutional authorization (Hauswald, 2025). Even in the absence of formal institutional endorsement, their judgments influence how students interpret correctness, misunderstanding, and progress. The core problem lies in the fact that AI systems exercise expert epistemic functions without being subject to the accountability structures that accompany human expertise. They do not operate within systems of certification, peer review, or appeal, yet their evaluative outputs shape learning decisions and cognitive trajectories.

In informal assessment environments, this situation leads to the structural exclusion of human educators. Educational experts lose the capacity to determine in advance how knowledge is organized within AI systems, as models derive their conceptual structures from training data without expert governance (Casper et al., 2024). They are also unable to audit the evaluative logic of AI systems, since the basis for specific judgments is distributed across non interpretable parameter spaces. When errors or systematic biases are identified, experts face significant obstacles to correction, as targeted intervention at the level of specific judgments is not technically feasible. These dynamics create a condition in which epistemic authority is exercised without institutional accountability, giving rise to significant risks within educational practice.

2.2 Human participation in public educational cognitive infrastructure

The challenges outlined above indicate that the central issue is not the technical performance of AI systems, but their emerging role within the epistemic environment of education. As AI systems increasingly mediate how students assess understanding and make learning decisions, they function as components of a broader cognitive structure. In this context, describing educational AI solely as a tool or platform fails to capture its systemic role.

This paper therefore adopts the concept of public educational cognitive infrastructure. In this usage, infrastructure refers to systems that exhibit widespread social dependency, long term integration into practice, and structural influence over possible forms of action and judgment (Sellar & Gulson, 2021). When AI systems are routinely used to organize learning content, evaluate understanding, and recommend learning pathways, they become embedded in the conditions under which educational activity takes place.

Such infrastructure possesses a distinct cognitive dimension. Educational AI systems do not simply provide access to information. They continuously participate in determining what counts as valid understanding and appropriate learning progression. These determinations involve judgments that are epistemic in nature and that have traditionally been associated with professionally accountable human roles within education.

For this reason, publicness constitutes a necessary feature of educational cognitive infrastructure. Publicness here does not refer to ownership, but to governance. Systems that influence educational judgment at scale cannot be governed exclusively by individual organizations without mechanisms for collective oversight. Decisions about standards of understanding and acceptable learning trajectories concern society's shared epistemic order and therefore require forms of participation that extend beyond private control.

Human participation within public educational cognitive infrastructure is essential to maintaining epistemic accountability. This participation does not imply that human experts must directly control every instance of AI mediated assessment. It requires the existence of interfaces through which cognitive structures can be made accessible to human interpretation, review, and modification. By externalizing aspects of knowledge organization and evaluative logic, such interfaces allow experts to contribute to guidance, auditing, and correction processes that are otherwise foreclosed.

Human participation thus functions as a condition for the legitimate operation of educational AI systems. It enables experts to engage with the structures that shape AI mediated judgment and preserves society's capacity to contest, revise, and collectively determine educational standards. Without such participation, educational AI risks consolidating epistemic authority in ways that are misaligned with the institutional foundations of education.

### 3. Technical foundations: the open cognitive graph (OCG)

3.1 From epistemic authority to a governance problem

In earlier sections, this paper argued that educational assessment derives its epistemic authority from institutionalized processes that support accountability, contestation, and revision. Pedagogical judgments are recognized as authoritative not because they are infallible, but because they are embedded in structures that allow expert review, appeal, and correction. When AI systems participate in educational evaluation and guidance, this authority is exercised without the institutional conditions that traditionally sustain it.

The technical foundations of contemporary AI systems help explain why this disjunction arises. In neural network based architectures, the assumptions that govern system behaviour, including how concepts are related and how learning progression is inferred, are not represented explicitly. Instead, they are distributed across large numbers of parameters learned through statistical optimization. Although these parameters collectively shape evaluative outcomes, they do not correspond to pedagogical concepts in a form that can be directly inspected, debated, or revised.

This creates a fundamental governance problem. Educational governance seeks to regulate pedagogical logic at the level where epistemic authority operates, namely at the level of conceptual structure, prerequisite relations, and standards of understanding. Neural network parameters, however, are accessible only at a numerical level that does not align with these pedagogical abstractions. As a result, there is a mismatch between what must be governed in order to ensure epistemic accountability and what is technically available for intervention.

The consequence is that epistemic authority is effectively exercised through system behaviour without being institutionally grounded. AI systems organize knowledge, infer learning needs, and evaluate understanding in ways that influence educational trajectories, yet the logic underlying these judgments remains opaque and insulated from established forms of professional oversight. Attempts to govern such systems by focusing on transparency at the parameter level or by regulating outputs alone do not address this structural gap. What is missing is an intermediate layer through which pedagogical authority can be articulated, examined, and contested.

3.2 Govern AI's epistemic authority through open cognitive graph

The OCG is proposed as a technical and representational response to the governance challenges discussed above (Li & Wang, 2025). Its purpose is not to render the internal mechanisms of neural networks fully interpretable at the level of parameters, but to provide a cognitive representation at the level of abstraction at which pedagogical reasoning actually operates, one that can be articulated, examined, and revised. By externalizing pedagogical assumptions that would otherwise remain implicit in model behavior into explicit structures, OCG creates the conditions for the re-institutionalization of epistemic authority.

Within OCG, pedagogical logic is expressed through explicit cognitive structures. Concepts are represented as domain-contextualized nodes, allowing the same term to be differentiated across disciplinary contexts. More importantly, the graph encodes structural pedagogical judgments through a set of relationship types with clear instructional significance. These relationships include prerequisite relations within a specified domain, in which one concept must be understood before another (prerequisite_of@Domain); structural analogical relations between concepts across domains that can facilitate transfer learning (analogous_to@D1↔D2); commonly recognized misconceptions in pedagogical practice that should be anticipated and addressed (common_misconception); and intermediate scaffolding concepts that bridge prior knowledge and target understanding (scaffolds). These relationships have traditionally relied on teachers' professional judgment and remained implicit, but are here rendered as explicit pedagogical structures that can be inspected.

By encoding these pedagogical relations as identifiable connections within the graph, the OCG shifts epistemic authority from an implicit effect of system outputs to a set of structural judgments that can be examined and debated. Authority no longer resides in the fact that a model produces a particular conclusion, but in the justification for why a given organization of knowledge, set of prerequisite assumptions, or learning pathway is considered appropriate. Because these judgments are represented explicitly, they can be reviewed by experts, points of disagreement can be made visible, and revisions can be undertaken in response to new evidence or pedagogical experience.

In addition, paths through the graph represent learning progression itself. The coexistence of multiple paths acknowledges pedagogical pluralism and avoids fixing a single learning sequence as the sole standard. Each concept and relationship is accompanied by provenance information indicating which experts contributed it, what evidence supports it, and when it was last validated. This provenance mechanism further anchors epistemic authority in concrete human judgment and professional responsibility.

The construction of OCG follows a human-in-the-loop process, ensuring that these pedagogical structures are not determined unilaterally by algorithms. AI systems assist by analyzing educational materials, expert demonstrations, and learner interaction data to propose candidate conceptual relations and learning pathways. Domain experts and experienced educators evaluate these proposals and, based on professional judgment, accept, reject, or modify them. When disagreements arise over prerequisite relations, analogical structures, or the identification of misconceptions, such disagreements are made explicit and addressed through deliberative mechanisms, rather than being systematically suppressed.

Through this process, epistemic authority is not transferred to the AI system but is re-institutionalized at the level of the OCG. The graph functions as a governable interface between AI systems and educational institutions, enabling human communities to retain ongoing participation in the adjudication of knowledge structures and learning pathways. In this way, OCG supports the operation of educational AI while restoring the alignment between epistemic authority and institutional accountability.

## 4. Governance architecture: the trunk-branch model

4.1 Epistemic authority as the object of governance

The governance of educational AI concerns not only technical reliability but the organization of epistemic authority. In established knowledge systems, authority over what counts as valid understanding is produced through collective and institutionalized processes. Disciplinary knowledge is validated through peer review, professional standards are negotiated through collective deliberation, and educational expectations are shaped through sustained interaction among practitioners. Within such arrangements, authoritative judgments remain open to challenge, reinterpretation, and revision.

Educational AI systems increasingly participate in functions that depend on these judgments. They organize conceptual structures, recommend learning sequences, and implicitly define standards of understanding. As these systems are adopted at scale, the cognitive structures they encode begin to shape shared educational practice. Epistemic authority thus becomes the primary object of governance, since it is this authority that determines how knowledge and learning are interpreted across contexts.

4.2 The limits of firm-level governance

When the governance of educational AI is concentrated within individual commercial entities, the social basis of epistemic authority is altered. Decisions about knowledge representation, pedagogical structure, and learning progression are made within organizational frameworks oriented toward product development and market strategy. These frameworks are not designed to function as arenas for collective epistemic deliberation.

Although firms may incorporate expert consultation or user feedback, final authority over cognitive structures remains centralized. The processes through which pedagogical assumptions are encoded are typically opaque to external scrutiny and insulated from sustained public contestation. As a result, epistemic judgments that would traditionally emerge from community level processes are relocated into private governance contexts.

This shift creates a structural mismatch. Corporate governance mechanisms are effective at managing risk, coordination, and accountability within firms, but they lack procedures for adjudicating contested epistemic claims or accommodating pluralism in educational values. As educational AI systems take on broader social roles, firm-level governance becomes insufficient for the scope of epistemic authority exercised.

## 4.3 Community governance as an architectural requirement

The need for community governance arises from the nature of epistemic authority itself. Because knowledge standards are socially produced and continuously revised, their governance requires institutions capable of supporting disagreement, revision, and collective judgment. Community governance does not presume consensus in all cases, but provides structured processes through which authoritative judgments remain contestable.

As educational AI systems encode cognitive structures that function as shared reference points, governance must operate at a level that allows communities of expertise and practice to participate in defining, reviewing, and revising those structures. Without such participation, epistemic authority risks becoming fixed through technical deployment rather than maintained through social processes.

The trunk–branch model is proposed as a governance architecture that responds to this requirement. It does not treat governance as an external constraint imposed on AI systems, but as an internal organizational principle that structures how epistemic authority is distributed, stabilized, and revised over time.

## 4.4 The trunk–branch model as a governance architecture

Within the trunk–branch model, epistemic authority is organized across differentiated layers. The trunk represents a consensus layer that encodes stable and well-established pedagogical structures. It provides a shared foundation that supports coherence and interoperability across educational contexts. Changes to the trunk require elevated justification and collective review, reflecting the role of consensus in sustaining epistemic stability.

Branches represent a pluralism layer built upon the trunk. They allow communities to adapt shared structures to local contexts, pedagogical traditions, and minority perspectives. Through branching, variation and experimentation are enabled without fragmenting the shared epistemic foundation. Successful innovations developed within branches can be proposed for integration into the trunk through established review processes.

This layered architecture ensures that epistemic authority is neither fully centralized nor fully fragmented. Stability and variation are coordinated through explicit structural relations, allowing educational AI systems to remain responsive to diverse contexts while preserving shared standards where appropriate.

4.5 Auditability and correction within the architecture

A governance architecture must include mechanisms for identifying and correcting errors. Within the trunk–branch model, auditability and correction are treated as structural features rather than ad hoc interventions. Logical consistency checks, educator feedback, learner outcome data, and research updates provide multiple channels through which potential problems are identified.

Errors are addressed through graduated escalation. Minor issues can be resolved within branches, while more significant problems are reviewed at the trunk level. In cases of critical failure, corrective mechanisms allow for rapid rollback to previously validated states. Corrections at the trunk level propagate to branches, ensuring that safety-critical updates are not overridden by local variation.

Through these mechanisms, the trunk–branch model translates community judgment into operational governance. Epistemic authority remains socially grounded, contestable, and revisable, while being expressed in forms that can effectively guide the behavior of educational AI systems.

## 5. Case study: A community-governed foundation model

5.1 Case setting and institutional preconditions

In earlier sections, this paper argued that educational AI can exercise epistemic authority in a legitimate manner only when it is embedded within community-based governance structures. To illustrate how this framework operates in practice, this section presents an envisioned case of a community-governed educational foundation model designed for introductory science education.

The model is operated by a non-profit foundation that holds it as a public trust. The foundation's charter explicitly defines educational purpose as the sole governing objective and prohibits commercialization or transfer of the model. This institutional arrangement ensures that the system is not oriented toward market returns or corporate strategy, but is instead constrained by public educational goals.

Governance authority is institutionally differentiated and allocated. An Academic Committee is responsible for maintaining the trunk layer, including the review of stable pedagogical structures and consensus-based knowledge. A Community Council represents branch maintainers and user constituencies, participating in the governance of plural pedagogical practices and localized educational needs. This dual governance structure establishes the institutional preconditions for subsequent technical and pedagogical decision-making.

5.2 Technical architecture and the source of authority

At the technical level, the model is not a standalone language model, but a system composed of multiple interdependent components (see Figure 1.). Its core architecture includes three elements: a neural language model that provides natural language interaction capabilities; OCG encodes the pedagogical structure of introductory science education; and inference mechanisms that constrain model behaviour according to the specifications of the OCG.

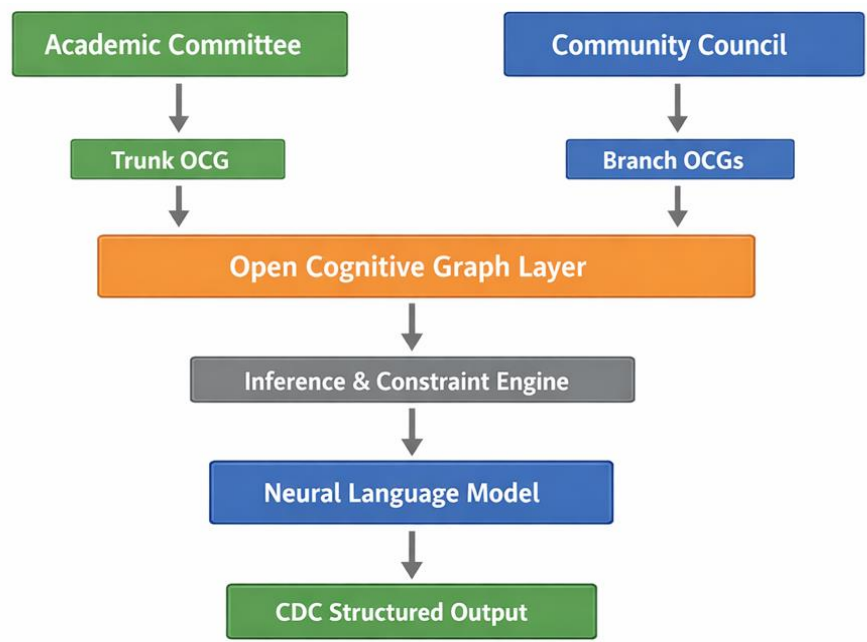

Figure 1. Governance-aware system architecture

Within this architecture, the OCG is explicitly designated as the authoritative source of pedagogical decision-making. The language model does not independently determine instructional structure or learning pathways. Instead, when generating explanations or guidance, it operates under constraints imposed by the pedagogical logic encoded in the graph. Decisions concerning whether prerequisite understanding has been established, which analogies are pedagogically appropriate, and which common misconceptions should be anticipated and addressed are determined by structural judgments represented in the OCG.

Through this design, pedagogical authority is no longer implicitly embedded in model parameters or emergent outputs. It is externalized and concentrated within a cognitive representation layer that is inspectable and revisable. This separation provides the technical foundation for governance, auditability, and correction.

To make such governance operational, the system can require that model outputs be expressed in a structured, auditable reasoning trace aligned with the OCG. One simple illustration is a "Concept–Domain–Claim (CDC)" response format, in which each step is labeled by its mathematical domain and the invoked theorem or rule:

**Illustrative CDC trace (right-triangle example)**

You are a CDC Math Assistant. Please answer in CDC format.

Problem: In a right triangle, the legs are a=3 and b=4. Find the hypotenuse c.

Requirements:

1) Show complete reasoning steps

2) Label each step with @math domain

3) State the theorem/justification for each step

Format:

Given --{Apply@Domain [Theorem]}--> Intermediate --{Compute@Domain}--> Final

Given: right triangle, a=3, b=4

--{Apply@Geometry [Pythagorean Theorem]}-->

c^2 = a^2 + b^2

Substitute values

--{Compute@Algebra}-->

c^2 = 3^2 + 4^2 = 9 + 16 = 25

Take square root

--{Compute@Algebra}-->

c = sqrt(25) = 5

Final answer: c = 5

In an OCG-governed system, the CDC trace is not merely a didactic style choice: it provides a stable interface for (i) constraining generation (only permitted theorem edges and domain transitions may be used), and (ii) validating outputs (each step can be checked against graph-encoded relations such as prerequisite concepts and admissible inference rules).

5.3 Problem emergence and community intervention

After the model is deployed in practice, specific pedagogical issues begin to surface. Educators using the system report that students consistently struggle with a key conceptual transition: moving from understanding "energy" as a property of objects to understanding "energy" as a conserved quantity within systems.

Further analysis reveals that this difficulty does not stem from deficiencies in the language model's expressive capacity, but from the pedagogical structure encoded in the OCG. The graph represents this transition as a direct prerequisite relationship, assuming that mastery of "energy as property" is sufficient for understanding "energy conservation." Experienced teachers, however, note that in classroom practice this transition typically requires intermediate scaffolding concepts, without which students struggle to make the necessary abstraction.

In this context, the community governance mechanism becomes operative. A consortium of middle school science teachers, acting as branch maintainers, proposes a modification to the graph structure based on instructional experience. The proposal introduces two intermediate concepts—"energy transfer" and "system boundaries"—between "energy as property" and "energy conservation," forming a learning pathway more consistent with observed learning processes.

5.4 Validation, decision-making, and structural repair

The proposed modification is not adopted directly, but enters an established validation process. Automated consistency checks first confirm that the change introduces no logical violations, such as circular prerequisite relations. Learning science researchers then evaluate the proposal for alignment with existing research on students' conceptual development of energy. In parallel, the modified structure is piloted in a limited setting, with student learning outcomes used to assess pedagogical effectiveness.

Once these validation steps yield positive results, the modification is formally adopted within the proposing branch. Based on accumulated empirical evidence and practical feedback, the change is subsequently submitted to the Academic Committee for consideration at the trunk level. After reviewing the documentation, the committee approves the modification to the trunk. The correction is then propagated to all branches through the trunk–branch mechanism.

Following the update to the OCG, the model's global behavior changes accordingly. All students now encounter the scaffolded conceptual pathway when engaging with relevant content. The modification is fully documented in the OCG changelog, including detailed provenance and validation information.

5.5 The feasibility of democratic fine-tuning

This case demonstrates that the governance architecture proposed in earlier sections is not merely normative in nature, but can be implemented through concrete technical mechanisms. Because the model's pedagogical decisions are constrained by the structure of the OCG, modifications to the graph translate directly into systematic changes in model behavior.

The process also illustrates how distributed expertise can be integrated into public educational infrastructure through institutionalized procedures. The problem is identified by practicing teachers, the solution is proposed within a community context, validated by researchers, and implemented through formal governance mechanisms. Epistemic authority is not concentrated in any single actor, but coordinated across roles with complementary forms of expertise.

Equally important, the process exhibits a high degree of transparency. The basis for system behavior is traceable, decisions can be examined and contested, and revisions can be made when warranted. Once a correction is validated, its benefits accrue to all users through shared infrastructure, rather than remaining confined to individual practices.

This case therefore suggests that, under appropriate governance and technical conditions, the fine-tuning of educational AI can function as a democratic, scalable, and institutionally legitimate process, rather than as a closed technical operation controlled by a single organization or model provider.

A common concern is whether such an approach is feasible given the cost of constructing and maintaining high-quality cognitive graphs. This paper does not assume that comprehensive graphs for all domains can be built upfront. Instead, the practical strategy is to begin with a minimal, foundational OCG (a "baseline trunk") that externalizes core concepts, prerequisite relations, and a small set of high-impact misconceptions as an auditable object. Even in this limited form, the graph already enables two operational functions demonstrated in our case logic: (i) constraining LLM outputs to follow an explicit pedagogical structure, and (ii) checking LLM outputs against graph-encoded relations for validation and error detection. Under this incremental approach, construction cost is staged and governance effort is focused where the educational risk and usage density are highest.

## 6. Implications

Framing educational AI as infrastructure fundamentally reshapes how issues of equity, policy, and sustainability are understood. Much of the current debate around educational technology equity focuses on access, emphasizing whether learners can reach AI tools or whether schools possess sufficient technical capacity to deploy them. An infrastructure perspective shifts attention away from access alone and toward the conditions under which AI-mediated cognitive environments are produced, governed, and stabilized. The central question becomes not merely who can use AI, but whether the pedagogical structures embedded in AI systems serve diverse learners in equitable and accountable ways.

High-quality pedagogical structure has historically been scarce and unevenly distributed. Deep expertise in how learning unfolds across concepts, how misconceptions arise, and how instructional scaffolding should be sequenced has tended to concentrate in privileged educational contexts, such as well-resourced schools or elite institutions. When pedagogical knowledge remains localized in individual practice, it functions as a private advantage rather than a shared resource. Encoding such expertise in OCGs and governing it as public educational infrastructure creates the possibility of transforming pedagogical knowledge into a public good. Learners and educators across contexts could benefit from collectively validated instructional structures that would otherwise remain inaccessible.

Equity also depends on who can participate in producing and exchanging pedagogical structure. For a broader educational community—including teachers, students, parents, tutors, and local organizations—the OCG can serve as a shareable medium for teaching and learning communication: educators can publish their own cognitive graphs (or branch variants) as reusable instructional pathways; learners can compare explanations across branches; and communities can organize collaborative review activities, workshops, and graph-building events to surface local needs and diversify pedagogical representations. In this sense, fairness is advanced not only by distributing access to AI tools, but by widening participation in the infrastructure that shapes what the tools teach.

This potential, however, is not realized automatically. Without deliberate governance, educational AI systems will reflect the pedagogical assumptions of those who design and train them. These assumptions are often shaped by elite educational environments and may be propagated as default or universal standards, even when they fail to align with the needs of diverse learner populations. An infrastructure approach makes this risk visible and highlights the necessity of community governance. Through structured participation, review, and revision, community governance enables multiple pedagogical traditions, cultural contexts, and learner needs to be represented within shared cognitive infrastructure, preventing the silent normalization of narrow instructional perspectives.

Viewing educational AI as infrastructure also carries significant implications for AI policy. Existing policy frameworks tend to focus on applications by regulating specific uses, requiring disclosures, or mandating human oversight at points of deployment. While such measures remain important, they do not address the deeper layer at which educational influence is exercised: the cognitive structures that shape understanding, progression, and evaluation across applications. An infrastructure-oriented policy approach directs attention to the governance of these underlying knowledge layers. This perspective supports the case for public investment in educational AI infrastructure oriented toward public purposes, much as governments invest in physical or informational infrastructure. It also motivates the development of interoperability standards, allowing OCG specifications to function across platforms and preventing lock-in to proprietary ecosystems. Certification requirements could further ensure that AI systems claiming educational deployment maintain auditable cognitive representations that meet defined governance criteria. Because educational infrastructure operates across national boundaries, international coordination becomes necessary to prevent fragmentation and to ensure broad participation in defining shared epistemic standards.

The infrastructure framework additionally reframes questions of sustainability for non-profit and community-governed AI systems. Such systems are often compared unfavourably to commercially funded alternatives in terms of computational scale or interface sophistication.

However, competition in educational contexts does not rest solely on technical performance. Community-governed systems compete on trust and epistemic accountability. Educational institutions increasingly face demands from parents, regulators, and professional communities to explain how AI systems make pedagogical decisions, who bears responsibility when errors occur, and whether underlying assumptions can be inspected and contested. Community-governed infrastructure is designed to provide clear answers to these questions, while proprietary systems typically cannot. In contexts where epistemic responsibility is integral to institutional legitimacy, this capacity for accountability becomes a durable competitive advantage. Sustainability, from an infrastructure perspective, thus rests not on scale alone, but on the ability to sustain trust through transparent and participatory governance.

## 7. Limitations

This study has several limitations. Firstly, treating educational AI as community-governed infrastructure introduces a set of structural tensions that must be acknowledged explicitly. One such tension concerns efficiency. Community governance is necessarily slower than unilateral corporate decision-making. Deliberative processes require time, consensus-building is demanding, and reconciling diverse perspectives involves negotiation and compromise. In technical domains characterized by rapid iteration and frequent capability advances, governance processes may lag behind model development and deployment. This tension is real and cannot be dismissed. At the same time, it is not unique to AI governance. Democratic societies routinely accept reduced efficiency as the cost of legitimate authority, particularly in domains where decisions have broad and lasting public consequences. The relevant question, therefore, is not whether community governance is slower, but whether decisions about educational AI should be treated as matters of product optimization or as matters of public epistemic policy. This framework adopts the latter position, while recognizing that doing so entails nontrivial efficiency costs.

A second limitation concerns conflict resolution. The trunk–branch model assumes that disagreements between branches and the trunk, or between community actors and academic committees, can be addressed through established deliberative procedures.

However, it is important to clarify that the trunk–branch structure is not intended to force the Academic Committee to "decide" all substantive controversies. In this framework, the trunk functions primarily as a curated storage layer for broadly shared axioms, well-established definitions, and consensus pedagogical dependencies that the field treats as stable. Where legitimate academic disagreement persists, or where choices are value-laden (e.g., cultural framing, curricular emphasis, or pedagogical philosophy), the appropriate response is not to collapse plurality into a single trunk decision, but to represent alternatives as explicit branches maintained by the relevant communities. The system therefore stores competing structures rather than adjudicating them as true or false, while still making their assumptions inspectable and governable. In practice, however, some conflicts may resist resolution. Communities may hold fundamentally incompatible views about pedagogical truth, disciplinary priorities, or educational values. In other cases, political or institutional pressures may distort technical judgment, undermining the ideal of evidence-based deliberation. These challenges do not admit simple solutions. Importantly, they are not unique to AI-mediated education. Traditional educational governance has long confronted analogous problems in curriculum design, textbook selection, and assessment standards, and has developed imperfect but functional mechanisms for managing persistent disagreement. The proposed governance architecture does not eliminate such conflicts, but situates them within transparent and structured processes, allowing AI governance to draw on existing educational precedents while adapting them to the specific technical characteristics of AI systems.

A third limitation concerns scope. Not all educational AI systems warrant infrastructure-level governance. Many tools, such as flashcard applications, gamified practice environments, or administrative support systems, do not exercise the form of epistemic authority that motivates the governance framework outlined in this paper. Distinguishing which systems cross the threshold into infrastructure, and therefore require OCG-based governance, is itself a matter of judgment and potential contestation. This paper has proposed structural mediation of knowledge and the exercise of epistemic authority as key criteria, but translating these concepts into operational thresholds will require further refinement through practice and policy debate. Scope determination should be treated as an ongoing governance question rather than a fixed classification.

## 8. Conclusion

Educational AI is increasingly shaping the cognitive environment in which learning takes place. Systems that mediate explanation, progression, and evaluation today will influence how future learners acquire knowledge, develop capabilities, and form understanding. This influence follows directly from current technological adoption patterns and institutional arrangements. When AI systems exercise such formative power, they function as infrastructure in a substantive sense, establishing foundational conditions for large-scale cognitive development. Treating these systems solely as commercial products or discretionary services risks structural consequences, including the concentration of epistemic authority, the entrenchment of narrow pedagogical assumptions, and the weakening of legitimate collective oversight over knowledge formation.

This paper has proposed an alternative approach: governing educational AI as Public Educational Cognitive Infrastructure. Central to this approach are OCGs, which externalize pedagogical structure in forms that can be inspected, audited, and revised, and governance mechanisms that anchor epistemic authority in community-based processes rather than proprietary control. The trunk–branch model provides an institutional architecture for coordinating stability and pluralism, while audit and correction mechanisms ensure ongoing accountability.

The framework advanced here is not presented as a normative ideal alone, but as a structural response to the role educational AI increasingly plays in society. If AI systems are to support learning in ways that remain responsive, contestable, and equitable, the cognitive structures they encode must be subject to democratic forms of governance comparable to those applied to other institutions that exercise authority over knowledge. In this sense, the governance of educational AI is inseparable from broader questions of public responsibility for the conditions under which understanding itself is produced.